%% file: ms.tex
\documentclass[compsoc,conference,a4paper,10pt,times]{IEEEtran}
\IEEEoverridecommandlockouts
\usepackage{cite}
\usepackage{amsmath,amssymb,amsfonts}
\usepackage{algorithmic}
\usepackage{graphicx}
\graphicspath{{figs/}}
\usepackage{textcomp}
\usepackage{bmpsize}
\usepackage{xcolor}
\usepackage{lipsum}
\usepackage{todonotes}
\presetkeys{todonotes}{inline}{}
\usepackage[colorlinks=true,urlcolor=black]{hyperref}
\def\BibTeX{{\rm B\kern-.05em{\sc i\kern-.025em b}\kern-.08em
    T\kern-.1667em\lower.7ex\hbox{E}\kern-.125emX}}

\newcommand*\circledw[1]{\tikz[baseline=(char.base)]{
    \node[shape=circle,draw,inner sep=.75pt, line width=.7pt] (char) {\scriptsize #1};}}

\begin{document}
\bstctlcite{IEEEexample:BSTcontrol}

\title{Resilient Self-Debugging Software Protection
\thanks{NB: appendices, if any, did not benefit from peer review.\newline A preprint of this paper has been deposited on ArXiv.\newline This research was funded
by the Fund for Scientific Research - Flanders (FWO) as part of project grant 3G0E2318; by the Flemish Government's Cybersecurity Initiative Flanders; and by
the TETRAMAX project, which has received funding from the European Union’s Horizon 2020 research and innovation programme under grant agreement number 761349.
}
}

\author{\IEEEauthorblockN{Bert Abrath, Bart Coppens, Ilja Nevolin, and Bjorn De Sutter}
\IEEEauthorblockA{\textit{Computer Systems Lab, Ghent University}, Belgium \\
bert.abrath@ugent.be, bart.coppens@ugent.be, ilja@nevolin.be, bjorn.desutter@ugent.be}
}

\maketitle

\begin{abstract}
  Debuggers are a popular reverse engineering and tampering tool. Self-debugging is an effective technique for applications to defend themselves against hostile
  debuggers. In penetration tests on state-of-the-art self-debugging, we observed several opportunities through which it could be attacked, however. We
  therefore improved upon the existing technique, making it more resilient by introducing reciprocal debugging and making the transfers of control between
  protected application and self-debugger more stealthy. This paper presents the improved self-debugging design, and details our research efforts into realizing
  reciprocal debugging. In our evaluation we show that the improved design is significantly harder for attackers to defeat. 
\end{abstract}


\input{introduction}
\input{attack_model}

\input{design}
\input{implementation}
\input{evaluation}
\input{conclusions}

\bibliographystyle{IEEEtran}
\bibliography{IEEEabrv,paper}

\end{document}

%% file: introduction.tex
\section{Introduction}

Debuggers are a popular tool in the toolbox of so-called man-at-the-end (MATE) attackers~\cite{collbergsurreptitious}. These attackers try to gain unauthorized
access to confidential assets in software by reverse engineering the software and try to tamper with the integrity of software assets by manipulating them,
e.g., to work around license checks or copy protections. Debuggers offer a convenient, interactive, scriptable, and flexible interface to inspect and tamper
with the evolving internal state of a running program.

Those same attacker goals can be reached with many other means available to MATE attackers that have white-box access to the software and the systems on which
they attack it in their labs. However, alternative tools such as emulators, instrumentation tools, and hypervisors lack the convenience of debuggers for many
concrete actions that attackers want to perform. Preventing the use of debuggers can hence slow down MATE attackers, and force them to invest more effort and
hence more money into their attacks. As the major objective of software protection is to change a MATE attack's (perceived) balance between required attack
investment and resulting profit~\cite{D4.01,D1.02}, techniques that prevent the use of hostile debuggers can hence be a worthwhile defensive investments.

One class of anti-debugging techniques has the application query the environment (e.g., via standard library APIs or system calls) for signs they are being
debugged or not~\cite{FerrisAntiDebugging2011,beginnersantidebugging,collbergsurreptitious}. However, as debuggers can intercept those queries, they are easy to
work around. For popular debuggers such as OllyDbg, a range of plugins exist to defeat such simple anti-debugging
techniques~\cite{scyllahide,titanhide,OllyExt}.

A stronger form of anti-debugging is by means of self-debugging~\cite{StarCraft2,nanomites}. The application then contains its own so-called self-debugger and
spawns its own debugging process, which attaches itself to the application as a debugger soon as it is launched. As all major OSs only support one debugger
process per debuggee~\cite{ptrace,mac_ptrace,russinovich2012windows}, the application-friendly self-debugger occupies the only seat available, such that a
hostile debugger can no longer be attached. The self-debugger is application-friendly, in the sense that it does not contain the traditional debugger
functionality that can be used to attack the application.

To prevent an attacker detaching the self-debugger to free the seat for their own hostile debugger, we previously proposed to migrate critical functionality
from the application to the self-debugger~\cite{abrath2016tightly}. Through source code annotations, a developer can designate the code to be migrated. During
execution, at certain program points control is transferred from the protected application to the self-debugger. The latter subsequently executes the migrated
(and transformed) fragments, and passes control back to the application. Simply detaching the self-debugger then breaks the program. Consequently, attacking a
self-debugging program to re-enable the use of a debugger requires much more work from the attacker, such as migrating the necessary functionality back into the
application (or into the dynamically linked library, which we can protect as well). The usefulness of such anti-debugging is apparent from the fact that our
approach is deployed in millions of devices protected by Nagravision, the company to which we transferred our technology.

Still, attackers have some opportunities to attack the self-debugging protection. In a public challenge and in professional penetration testing
experiments~\cite{D4.06}, we observed two critical aspects. First and foremost, if an attacker can still debug the self-debugger, they can observe the migrated
functionality and the control transfers. This can help them to work around the protection by undoing it or by circumventing it in some way. Secondly, when the
code fragments implementing the transitions and the interface to pass data between the application and the self-debugger lack stealth, attackers can use
relatively simple static techniques such as pattern matching to identify them, easing the implementation of (semi-)automated workarounds. We observed both
attacks in the aforementioned penetration testing experiments.

To counter those activities, we studied extensions in two directions: more stealthy control flow transfers that are also more resilient to workarounds on the
one hand, and reciprocal debugging on the other hand, because when both the protected application and the self-debugger debug each other, an attacker can debug
neither of them.

This paper reports our findings. Both for making control transfers more stealthy and resilient, and for making self-debugging reciprocal, we report
opportunities and obstacles as present on popular architectures and operating systems. We make the following contributions:

\begin{itemize}

\item We present an improved design of our self-debugging technique, featuring reciprocal debugging and stealthier interfaces.

\item We describe a working reciprocal debugging implementation, and explain the deadlocks inherent in such a system and how to handle them.

\item We examine portable methods through which a debugger can be stealthily signaled.

\item We provide a security evaluation of our improved design, using Binary Ninja.

\item We provide an open-source prototype implementation of our design, to be released at a later date.

\end{itemize}


%% file: attack_model.tex
\section{Self-Debugging Security Analysis}
\label{sec:attack_model}

\begin{figure*}
  \centering
  \includegraphics{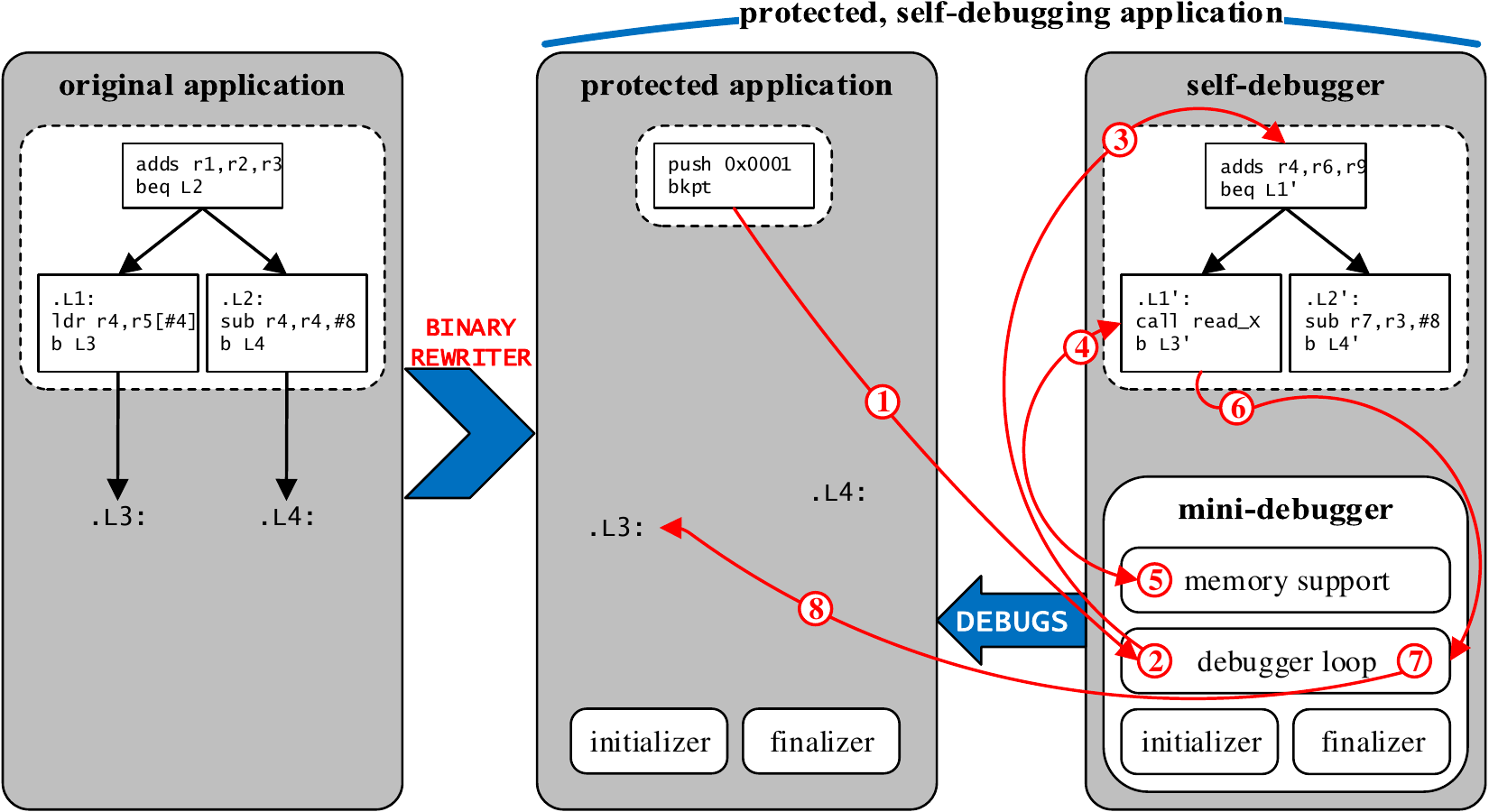}
  \caption{Overview of the original self-debugging protection}
  \label{fig:overview_original}
\end{figure*}

MATE attackers want to analyze or tamper with the sensitive assets contained in a target application, and can be assumed to have total control over the
environment this application runs in~\cite{emse2019}. This means they can run the application in an emulated environment, disable security features provided by
the OS such as ASLR, as well as use many other different tools at their disposal to analyze or tamper with an application statically or dynamically. Debuggers
are powerful dynamic tools, providing attackers with the capabilities to gather traces, dump cryptographic keys, or even execute code fragments out of context,
e.g., to work around copy protections.

All major OSs only provide support for a single debugger per application~\cite{ptrace,mac_ptrace,russinovich2012windows}. Self-debugging takes advantage of this
limitation by occupying the only available debugger seat with a custom, application-friendly self-debugger~\cite{abrath2016tightly}. This protection prevents
attackers from attaching their own, hostile debugger to the application, and consequently first has to be defeated by any attackers before they can debug the
application. The protection has thus become the target, and we therefore investigate how it can be attacked as well as defended.

We start with an overview of our original design, and then explore the different ways in which it can be attacked.

\subsection{Self-Debugging Overview}

Figure~\ref{fig:overview_original} illustrates the design of the original self-debugging technique~\cite{abrath2016tightly}. Whereas the original, unprotected
application as shown on the left consists of a single process, the protected self-debugging application, which was generated by means of binary rewriting,
consists of two processes. The first process is the \emph{protected application}, which still contains the main part of the application code, largely unchanged.
The second process is the \emph{self-debugger}, which provides the protection to the application by ``debugging'' it. In the context of this paper, this means
``attaching to the application like a debugger does'', it does not include the act of using debugger functionality to find bugs. The self-debugger contains a
\emph{mini-debugger} component that implements the necessary debugger interface functionality using \texttt{ptrace}, the standard debugging API on
Linux~\cite{ptrace}. Next to that it also contains a number of \emph{code fragments} migrated from the application context, guided by source code annotations.
These fragments get invoked by the protected application whenever they need to be executed, but they will be executed in the self-debugger's context. It is this
mechanism that creates a dependence between the protected application and the self-debugger. This dependence prevents the execution of the application when the
self-debugger is not attached, and thus keeps an attacker from detaching the self-debugger and attaching their own hostile debugger instead.

The initialization of the self-debugging technique is provided by an initialization routine that is injected into to the binary. This routine is invoked
whenever the binary is loaded, both in case the binary is an executable file with which a process is launched and in case it is a dynamically linked library
loaded into an existing process. The routine performs a \texttt{fork}~\cite{fork}, with the parent becoming---or rather, staying---the protected application,
and the child becoming the self-debugger that attaches to the application. The application thread that invokes the initialization routine is kept waiting and
only allowed to continue after the self-debugger has finished attaching to all of the protected application's threads. There is also an accompanying
finalization routine.

At run time, the protected application needs to execute migrated code fragments. The red edges in Figure~\ref{fig:overview_original} show the flow of control to
invoke and execute such a fragment in the self-debugger. The application contains exception-inducing instructions at every location a fragment is to be invoked.
In our original implementation, these exception-inducing instructions are hard-coded breakpoint (\texttt{BKPT}) instructions preceded by instructions that push
an identifier onto the stack. When they get executed an exception occurs that generates a signal. The kernel then stops the protected application until its
debugger investigates the signal and allows the application to continue. The exception thus switches control flow to the self-debugger, and more specifically to
its mini-debugger component, which derives the requested code fragment from the identifier on the stack: It uses this identifier to look up the associated
fragment's address in a mapping table, and then transfers control to that address. The migrated fragment subsequently executes in the self-debugger context, but
performs reads and writes to the protected application's address space when required, using the mini-debugger's memory support. Upon return from the fragment,
the mini-debugger updates changed registers in the application process, and finally allows it to continue executing. The mini-debugger itself then continues in
its debugger loop, awaiting further requests by the application.

\subsection{Defeating Self-Debugging}

To attach their own hostile debugger functionality to the protected application to observe or manipulate its execution, attackers first have to defeat the
self-debugger. They can achieve this in a number of ways. We present these attacks in two broad categories: those where the self-debugger is detached, and those
where it is still attached. In the first category a hostile debugger can be attached directly to the protected application. In the second category, the hostile
debugger cannot be attached directly.

\begin{figure*}[t!]
  \centering
  \includegraphics{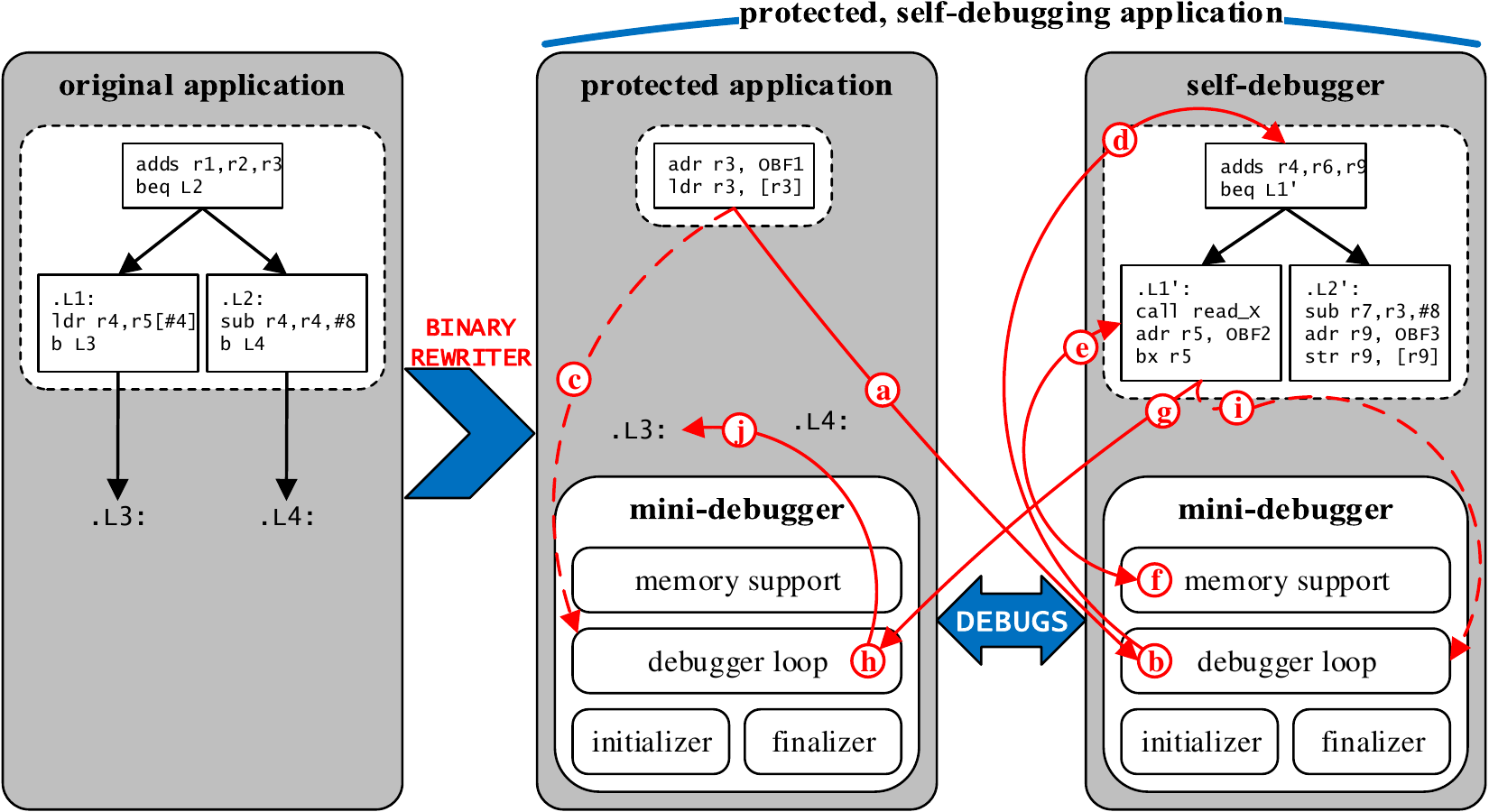}
  \caption{Overview of the improved self-debugging protection featuring reciprocal debugging and stealthier interfaces}
  \label{fig:overview_improved}
\end{figure*}

\subsubsection{With the Self-Debugger Detached}

Detaching the self-debugger is not hard by itself. It can be done, e.g., by killing the self-debugger, or by removing or disabling the initialization routine
that spawns it. This by itself does not suffice, however, because the protection also includes the migration of code fragments into the self-debugger. As the
protected application at times needs to execute these code fragments, it will stop working when the self-debugger is detached. Attackers thus need to deal with
these migrated code fragments. They can do so in several ways: The migration transformations can be reversed, the execution of fragments can be avoided, and
they can be emulated in an attacker-controlled process.

Reversing the transformations means both undoing the changes to the migrated code fragments so they can execute directly in the application again, and reversing
the control flow switches to the mini-debugger back to normal control flow that does not depend on a debugger. To reverse transformations in general, attackers
first need to find out which transformations have been applied and where, after which they can tamper with the binary to undo the protection. For
self-debugging, attackers need to locate the control flow switches and the migrated code fragments. Simple static analysis suffices to do so: one can directly
find the hard-coded breakpoints by simply scanning the binary, and one can find their intended targets by using the accompanying identifier to look up the
target address in the mapping table. Alternatively, dynamic analysis such as tracing the self-debugger's execution can also be used to determine the intended
targets of the control flow switches. Such an analysis can be achieved easily by attaching a hostile debugger to the self-debugger process. In the public
challenge of the ASPIRE FP7 project, the self-debugging protection was reversed manually by an amateur hacker in this way~\cite{D4.06,emse2019}. With the
necessary effort, attackers could automate these analysis steps, as well as the subsequent tampering.

Alternatively, attackers might be interested in tracing or tampering with only a small part of the application, in which case reversing all the transformations
for all of the migrated code fragments is typically not necessary. The attackers can instead try to invoke execution paths that enable them to reach their
specific goals, yet contain no (or few) migrated code fragments. Using a debugger to alter the flow of execution at will---perhaps even in a scripted
manner---is of great benefit in constructing such paths. Moreover, when the code to be protected is a dynamically linked library, attackers can invoke exposed
API functions in the order of their choosing. This attack approach was employed by professional penetration testers during the ASPIRE FP7 project's security
evaluation~\cite{D4.06,emse2019}. The same analyses as used for reversing transformations are required, as well as the extra path analysis.

A last attack option is to leave all transformations intact, and to emulate the mini-debugger function. This emulation comes down to switching control flow to
the right migrated code fragments when requested, and creating the right context for them to execute. One way to do so is to create and attach a hostile
debugger that contains the necessary emulation functionality next to the traditional debugging functionality one expects to find in any typical debugger. The
migrated code fragments can then be executed in this hostile debugger's context, while it can use its privileges to tamper with and trace the protected
application at will. Another option would be to insert some kind of signal handler that can resolve the control flow switches into the application itself. In
that scenario, however, the migrated code fragments would run directly in the target's address space. As the fragments were transformed to run in another
address space, some hooking of the \texttt{ptrace} API would be required for these fragments to execute correctly.

\subsubsection{With the Self-Debugger Attached}

Here, attackers cannot attach their own debugger to observe or to tamper directly with the protected application. However, they can target the self-debugger
itself, or gain debugging privileges through some other way than the \texttt{ptrace} API. During the ASPIRE Security Evaluation professional penetration testers
tried and failed to use multiple tools for tracing or live-debugging a protected application~\cite{D4.06}.

Attackers can attach their debugger directly to the self-debugger, and use its privileges to subvert the protection, effectively turning the self-debugger into
their own hostile debugger. By debugging the self-debugger they have direct access to the migrated code fragments during their execution. More importantly, they
can also gain indirect access to the protected application by taking advantage of the self-debugger's privileges and injecting \texttt{ptrace} requests. This
allows them to trace and tamper with the application, without directly debugging it or interfering in its workings. Such an approach would require little
analysis, yet bypass the protection completely.

Alternatively, debugging-like powers can also be achieved without using the \texttt{ptrace} API. Program emulation using Valgrind~\cite{valgrind}, or
QEMU~\cite{bellard2005qemu} can provide tracing, as long as the \texttt{ptrace} API and the entire workings of the control flow switches are correctly emulated.
Using Valgrind, interactive debugging can even be provided through a gdbserver. We tried attacking our original implementation using this approach, but did not
succeed because of incorrect emulation of the \texttt{ptrace} API and the exception-inducing instructions~\cite{abrath2016tightly}. Implementing a correct
emulation of these aspects is just engineering, however. We therefore consider it worthwhile to defend the application against being executed in an emulated
environment. Many such defenses already exist, and are orthogonal to our technique~\cite{paleari2009fistful}. We thus consider anti-emulation out of scope for
this paper.

Finally, attackers could modify the OS in order to provide them an alternative to the \texttt{ptrace} API. This would come at the cost of considerable
engineering effort, however. To be useful to attackers, such an API has to provide the functionality necessary for interactive debugging or performing scripted
attacks.


%% file: design.tex
\section{Resilient Self-Debugging Design}
\label{sec:design}

To counter the above attacks, our improved self-debugging technique features two improvements: a tighter, reciprocal coupling between the application process
and its self-debugger process, and stealthier interfaces for invoking the mini-debugger. Figure~\ref{fig:overview_improved} contains an overview of the design.
We discuss many aspects below.

\subsection{Reciprocal Coupling}

The self-debugger can be a target for attackers, who either want to debug it or run the protected application without it being attached. We therefore explored
methods of making the coupling between the two tighter, that is, making it harder for an attacker to kill, detach, or subvert the self-debugger without
triggering a reaction. More specifically, we explored reciprocal debugging, where not only does the self-debugger ``debug'' the application, but the application
itself also ``debugs'' the self-debugger. This keeps any hostile debuggers from attaching to either. Figure~\ref{fig:overview_improved} presents an overview of
reciprocal debugging, together with the changes to the original design it requires in terms of components and the flow of control. These changes serve to ensure
that the protected application remains attached to the self-debugger. In the original design, the protected application could raise exceptions to invoke the
mini-debugger. In the improved design, such invocations can happen in two directions. Consequently, both processes contain an identical mini-debugger.

With two mini-debuggers present and the accompanying bidirectional mini-debugger invocations, the interactions between the protected application and the
self-debugger are more complex. We start with an abstract overview of these interactions, and follow up by illustrating the control flow portrayed in
Figure~\ref{fig:overview_improved} as a concrete example. At any time during the execution of application code, either the protected application or the
self-debugger is functioning as the \emph{exception-throwing process}, with the other process functioning as the \emph{exception-catching process}. The
exception-throwing process executes (possibly migrated) application code, until it throws an exception to invoke the mini-debugger. The exception-catching
process, on the other hand, executes in the debugger loop of its mini-debugger, awaiting any invocations. We consider the exception-catching process to contain
an \emph{active} mini-debugger. When an exception is thrown by the exception-throwing process, this process is stopped, and the active mini-debugger in the
exception-catching process is invoked. At this point, the roles of the two processes have to switch. The exception-catching process will transfer control to
application code and become the exception-throwing process. Conversely, the exception-throwing process will ready itself for invocations by activating its
mini-debugger, and thus become the exception-catching process. These transitions are managed by the active mini-debugger in the exception-catching process. It
first uses its debugger privileges to intervene in the stopped exception-throwing process, transitioning it to an exception-catching process. More precisely,
the active mini-debugger manipulates the state and context of the exception-throwing process and then resumes it in the debugger loop, thereby activating its
mini-debugger. After this transition, the active, invoked mini-debugger can complete the role switch by transferring control to the requested application code,
thereby deactivating itself and transitioning its process from exception-catching to exception-throwing.

We present the control flow depicted by the alphabetically ordered red edges in Figure~\ref{fig:overview_improved}) as a concrete example. Here, the protected
application is the exception-throwing process, executing application code; the self-debugger is the exception-catching process, with an active mini-debugger.
The protected application throws an exception (\circledw{a}), which invokes the active mini-debugger running in the self-debugger (\circledw{b}). First, the
active mini-debugger transitions the protected application to an exception-catching process by activating its mini-debugger (\circledw{c}). Secondly, the active
mini-debugger transitions the self-debugger to an exception-throwing process by transferring control to application code, in the form of a migrated code
fragment (\circledw{d}). This migrated code might perform memory operations, requiring the memory support of the mini-debugger (\circledw{e} and \circledw{f}).
At some point the migrated code in the self-debugger throws an exception (\circledw{g}), invoking the active mini-debugger running in the protected application
(\circledw{h}). This active mini-debugger also manages two transitions: First, of the self-debugger into an exception-catching process (\circledw{i}); Second,
of the protected application into an exception-throwing process, by transferring control to application code (\circledw{j}).

As an additional protective measure, the protected application and self-debugger attach to each other with the \verb!PTRACE_O_EXITKILL! option enabled. This
option ensures that a debugged process can never escape its debugger's control, even if the debugger unexpectedly exits. The kernel enforces this by immediately
shutting down the debuggee when a debugger exits. If the debugger simply detaches, however, the debuggee is allowed to continue normally. In the case of
reciprocal debugging, enabling this option ensures an attacker can kill neither the protected application nor the self-debugger, without the other immediately
shutting down.

If the entirety of its initialization happens correctly, this design is impervious to dynamic tampering through a debugger. Once the initialization is done, the
protected application can no longer be debugged, the self-debugger cannot be debugged, and even killing the self-debugger will not help an attacker.

\subsection{Stealthier Interfaces}

Regardless of the specific attack, analyzing how the mini-debugger is invoked and to which (migrated) application code it transfers control helps the attacker
in understanding the application. In our original design, the interfaces between the protected application and mini-debugger were implemented with manifest
idioms (i.e., breakpoint instructions and IDs pushed onto the stack), and the control flow between them was hence easy to follow, e.g., in a static analysis
that is customized to handle the idioms. To inhibit analysis, we extended the mini-debugger with stealthier interfaces.

This interface consists of two aspects, as the exception-throwing process (1) needs to invoke the mini-debugger in the exception-catching process by raising a
signal, and (2) needs to pass the necessary information to identify the target code to be executed once control is transferred. Whereas our original design only
allowed for a single implementation of both aspects, we explored and implemented several stealthier, alternative methods in our improved design for both of
them.

Moreover, the identification of target code fragments is also more flexible in the new design, as control can now be passed to any code fragment in the
exception-catching process, whichever of the two processes is the exception-catching one at some point in time. In the original design by contrast, control
could only be passed to the set of migrated code fragments. Because the identification of target code is more flexible, the interface through which both the
mini-debuggers in both the processes are invoked can be exactly the same for invocations in both directions of the reciprocal design. This is a useful feature
for two reasons. First, one mini-debugger implementation then suffices for both processes, so only one mini-debugger needs to be injected into the protected
binary. Secondly, having the same interface in both directions implies that the static code fragments making invocations do not reveal which mini-debugger
they invoke by means of their appearance alone. This reduces the amount of useful information that attackers can extract from the protected binary through
simple static attacks such as pattern matching.

\subsubsection{Signaling the Mini-Debugger}

We replaced the conspicuous \texttt{BKPT} instructions of our original design with injected code that looks harmless and similar to the surrounding code, yet
intentionally generates software faults. Whereas originally a \texttt{SIGTRAP} signal was generated and examined by the mini-debugger, our new design can handle
a mix of different signals such as \texttt{SIGSEGV} and \texttt{SIGFPE}. Examples of injected code signaling the mini-debugger from both the application and
self-debugger processes can be observed in Figure~\ref{fig:overview_improved} (\circledw{a} and \circledw{g}). Here, \texttt{ADR} is an ARM assembler
macro-instruction that produces an address in a register. The values \texttt{OBF1}, \texttt{OBF2}, and \texttt{OBF3} produced by the \texttt{ADR} instructions
are addresses that cause segmentation faults when accessed, either because they are invalid or because more permissions are required. We thus have some faulting
instructions (\texttt{LDR}, \texttt{BX}, and \texttt{STR} in Figure~\ref{fig:overview_improved}), preceded by some other instruction(s) that set up the fault
(the \texttt{ADR} instructions in Figure~\ref{fig:overview_improved}).

Figure~\ref{fig:overview_improved} depicts the injection of both the instructions that set the registers to set up the fault and the faulting instruction
itself, such that they directly follow each other. Our implementation has more complex features to inhibit analyses. Instead of injecting faulting instructions,
it can reuse instructions already present in the application. It then injects instructions that set up the registers for the software fault and add a jump to an
already existing fragment that ends with an exception-causing instruction (e.g., a \texttt{BX} or a \texttt{STR}) when executed on the set register contents.
This reuse makes the attacker's analyses harder as some instructions originally executed in only one context (the original application) are now executed in
multiple contexts, i.e., both as parts of the original application code and to generate faults. It also splits the code that generates the software fault from
the code that actually wants to invoke the mini-debugger, separating them in space. It even separates the software fault and its set-up in time when the
jumped-to fragment contains instructions before the faulting instruction. Proper care then has to be taken that the intervening instructions do not change any
live registers or have other side effects, of course.

By using less conspicuous code for signaling, however, it also becomes harder for the mini-debugger to distinguish between ``forged'' signals that request the
switch to application code, and signals generated by genuine software faults. Causes of such signals can be bugs in the application, but also intentional
functionality, such as processor feature detection, i.e., testing whether a certain instruction is supported by the hardware. Consequently, it is necessary
for the mini-debugger to correctly make this distinction, and it thus contains a mechanism that allows for the verification of requested switches.

\subsubsection{Target Identification}

There are two aspects of target identification to consider: the location in which to pass the information about the target, and the way in which this
information encodes the identity of the target. In our original design, both of these aspects were problematic with respect to stealthiness and ease of reverse
engineering. First, the mapping table used for converting identifiers into addresses was a single centralized and unobfuscated table of all the addresses of
migrated code fragments. A reverse engineer could thus, upon finding this table, find all control flow targets and start analyzing them. Secondly, in our
original design the identifier was passed in a fixed place---the top of the stack---by the instruction immediately preceding the trap.

In our improved design, we take a decentralized approach: rather than passing an identifier that needs to be linked to a target somehow, we now pass the address
of the requested target itself, in an obfuscated form. This way, there is no single table that can be abused by attackers to find all control flow targets.
Moreover, the obfuscated address is no longer passed in a fixed location, but in a variable location that is implicated in the software fault used to signal the
mini-debugger. This approach can be observed in Figure~\ref{fig:overview_improved}. In one instance, the code fragment in the self-debugger signals the
mini-debugger by jumping to the address \texttt{OBF2} (\circledw{g}), for which the execution memory permission is lacking. The mini-debugger finds this
obfuscated address \texttt{OBF2} by inspecting the software fault, and then decodes it into the target address to be invoked. Another example would be a a
software fault caused by a division by zero, where the obfuscated address could be stored in the register that was used as the numerator.

In our current proof-of-concept implementation we deploy only a few methods to obfuscate the target addresses. This could easily be extended to using a myriad
of data obfuscation methods. For every invocation the mini-debugger would then employ a decision tree to determine the specific method to decode the obfuscated
address. The variables used in the decision tree could include the signaling method, the values in fixed or method-dependent memory addresses or registers, etc.


%% file: implementation.tex
\section{Implementation Challenges and Solutions}
\label{sec:implementation}

This section discusses noteworthy aspects of our implementation that featured significant research challenges. In the process of exploring a more capable
mini-debugger, we came to rely on more recent \texttt{ptrace} functionality such as \verb!PTRACE_SEIZE! and \verb!PTRACE_O_EXITKILL!~\cite{ptrace}. These
features are only present as of Linux 3.4 and 3.8, respectively. Our binary rewriter operates on ARMv7 binary code.

\subsection{Reciprocal Debugging without Deadlocks}

A system of two processes that debug each other and communicate via signals is fragile and prone to deadlocks. To make it work correctly took significant
research efforts.

When a signal other than \texttt{SIGKILL} arrives at a debuggee process, the kernel places this process in a stopped state~\cite{ptrace}. The debugger is
notified of the signal, handles it (however it chooses to do so), and decides whether or not it is to be delivered to the debuggee. The debugger can thus choose
to suppress signals, which in effect the mini-debugger does when handling a request to transfer control to some application code. After the debugger is done
handling the signal, it allows the debuggee to continue (with or without actual delivery of the signal). In essence, when a signal arrives at a debuggee
process, it is stopped, and can only continue after its debugger allows it to. In the context of reciprocal debugging, this can become a problem. When signals
arrive at both the protected application and the self-debugger simultaneously or close together, both processes end up in a stopped state, each waiting until
the other one allows it to continue.

Conceptually, avoiding such deadlocks is simple: One can prevent signals arriving close together to both processes by preventing signals from arriving at the
exception-catching process altogether. That process will then at all times be available to handle the signals arriving at the exception-throwing process. This
is easier said than done, however. Avoiding deadlocks requires recognizing all possible sources of signals, and implementing methods to handle these cases. We
describe several specific sources of signals, the associated deadlocks, and the solutions. We conclude by discussing the remaining possible deadlocks.

\subsubsection{Child Processes}

When something happens to a child process of potential interest to the parent process---such as the child process stopping or exiting---, the kernel by default
sends a \texttt{SIGCHLD} signal to that parent~\cite{wait}. In addition, the kernel always treats a debugger process as a parent process of the
debuggee~\cite{ptrace}, so whenever a debuggee changes state, its debugger is notified through a \texttt{SIGCHLD} signal. In our case, any state change in any
of the two processes results in a \texttt{SIGCHLD} signal being sent to the other process. When the exception-throwing process invokes the active mini-debugger,
this could cause a deadlock. Without precautions, the exception-throwing process is placed in a stopped state by the kernel, which simultaneously sends a
\texttt{SIGCHLD} signal to the exception-catching process, which will be blocked until the exception-throwing process allows it to continue. This is precisely
the sort of situation we need to avoid.

To do so, we let both the application and the self-debugger process make use of the \texttt{sigaction} API~\cite{sigaction} to disable the default sending of
\texttt{SIGCHLD} signals by the kernel upon state changes in their child processes.

For applications where receiving and handling such signals from other child processes is part of the application's correct functioning, simply disabling the
signals is not an option. An alternative solution, which we have not implemented yet but only requires some engineering, is then to keep these signals pending
using the \texttt{sigprocmask} API~\cite{sigprocmask}, and only handle them at specified synchronization moments, when it is certain the self-debugger's process
state will not change.

There is one exception to the ignoring of \texttt{SIGCHLD} signals: When a debugged process exits, the kernel always sends one, which cannot be disabled.
However, before a debugged process is allowed to exit, its debugger is notified of the impending departure. The debugger can use this opportunity to detach from
the exiting process, forestalling the arrival of the \texttt{SIGCHLD} signal. We hence solved this exception by making the self-debugger detach from the
protected application in the appropriate places.

\subsubsection{Users and IPC}
\label{sec:users}

A user or process can send any kind of signal directly at a process using the \texttt{kill} command or API, provided that they possess the required
privileges~\cite{kill}. Such signals might be used to request that the application output a status update, or in the context of inter-process communication
(IPC). While we might question the motives of a human user simultaneously sending a signal to both the protected application and the self-debugger, there is at
least one system process that has a good motive. When a system shutdown is requested, the system daemon---be it \texttt{sysvinit}, \texttt{systemd}, or
something else---handles this by sending a \texttt{SIGTERM} signal to all running processes at the same time~\cite{shutdown}. Most processes will then
terminate, and any process that has not yet done so after a specified delay will be sent an additional \texttt{SIGKILL}, immediately terminating it. For
reciprocal debugging, the initial \texttt{SIGTERM} causes a deadlock that not even the subsequent \texttt{SIGKILL} can solve.

Using the aforementioned \texttt{sigaction} API does not suffice now: When a debuggee process asks not to be notified of signals from a specific type, the
kernel still informs its debugger~\cite{sigaction,ptrace}, which then first needs to intervene. In our case, we hence still have a deadlock. A working
alternative is blocking the sending of signals of a certain type with the \texttt{sigprocmask} API~\cite{sigprocmask}. Those signal are then kept pending until
the type is unblocked again. For a debugged process, only when the signal is unblocked again will the debugger be notified of its arrival.

Whenever a mini-debugger becomes the active one, it therefore blocks all possible signal types except for \texttt{SIGKILL} and \texttt{SIGSTOP}, which we
discuss later. This blocking is undone right before deactivating and transferring control to application code. The exception-catching process thus blocks all
possible signals, while the exception-throwing process maintains its original behavior. This maintains behavior that depends on signals sent by other processes
or users. The previously discussed case of a system shutdown serves as an example. There, the exception-throwing and the exception-catching processes
simultaneously receive a \texttt{SIGTERM} signal. The signal arriving at the exception-catching process is simply blocked. The signal arriving at the
exception-throwing process is first presented to the exception-catching process, which sends it on to the former. The application then shuts down in a proper
manner, cleaning up properly before exiting.

\subsubsection{Software Faults}

The aforementioned solutions only block signals that were sent to it through \texttt{kill} or one of its variants. A signal caused by a software fault inside
the process itself cannot be blocked that way, and has to be handled before the process can continue.

Faults inside the exception-throwing process can be intentional, i.e., injected by us to invoke the mini-debugger, or genuine faults, possibly caused by bugs
or, e.g., by code testing the processor's support for certain instruction set extensions. In Section~\ref{sec:switch}, we describe an optional mechanism to
distinguish between these cases and to pass any genuine signal on to exception-throwing process. It is thus no problem if that process generates a genuine
software fault. On the other hand, if the exception-catching process generates such a fault, a deadlock is certain to follow. A software fault generated in the
exception-catching process can only be the result of a bug in the mini-debugger, so with proper engineering this becomes a non-issue.

\subsubsection{Remaining Deadlocks}

An important caveat to the above is that neither \texttt{SIGKILL} nor \texttt{SIGSTOP} can be ignored or blocked, as described in Section~\ref{sec:users}. If a
\texttt{SIGKILL} arrives at either process, it simply dies, taking the other process with it because of the \verb!PTRACE_O_EXITKILL! option. This is not a
problem, it is exactly what we want. If a \texttt{SIGSTOP} arrives at the exception-catching process, a deadlock does still occur, as the exception-throwing
process is running application code instead of paying attention to its debuggee. This means that no \texttt{SIGSTOP} should ever be sent to the protected
application or the self-debugger, as either could be the exception-catching process at any moment.

This is certainly a limitation, but is not necessarily an important one for the specific applications we want to protect. On the one hand, an application
sending a \texttt{SIGSTOP} signal to stop itself or one of its component processes can be adapted to send a \texttt{SIGTSTP} signal instead. Contrary to a
\texttt{SIGSTOP}, a \texttt{SIGTSTP} can be blocked. Although the resulting process stop might be delayed, no deadlock will occur. On the other hand,
\texttt{SIGSTOP} signals originating outside of the application are entirely out of our control. Such signals are of use to potential attackers that want to to
tamper with the application, however, and the resulting deadlock could hence be seen as extra protection.

\subsection{Signaling the Mini-Debugger}

We use intentional processor faults to signal the mini-debugger. Such faults can be generated in various ways, but ultimately result in just four different
signals: \texttt{SIGILL}, \texttt{SIGFPE}, \texttt{SIGSEGV}, and \texttt{SIGBUS}. Popov et al.\ used such intentional faults for static control flow
obfuscation~\cite{popov2007binary}. Our context differs in two ways. First, whereas they aim to thwart disassembly, we aim to thwart hostile debuggers. Second,
we use the ARMv7 architecture, which is used on a variety of platforms with many optional features. Portability is thus a major concern. We investigated the
portability of various signaling methods for the ARMv7 architecture, and also compared with x86 (or x86-64) architectures to provide a broader scope.

\subsubsection{Illegal Instructions}

Not all ARM or x86 processors implement all extensions of their respective instruction set architectures (ISAs). Moreover, in both ISAs many combinations of
bits do not encode an actual instruction. For any processor, unsupported encodings are illegal instructions, and attempting to execute them results in a fault.
On Linux, such a fault generates a \texttt{SIGILL} signal.

Inserting an illegal instruction encoding is not future-proof, however, as future ISA revisions could redefine such encodings. Unimplemented instructions differ
between processors, and are hence not portable. Furthermore, instances are easy to identify using a disassembler.

\subsubsection{Division by Zero}

A less conspicuous approach uses instructions that only fault under some conditions, but otherwise execute normally. A good example are divisions, which only
fault for division by zero. Using data obfuscation techniques, loading a zero into the denominator register can remain hidden from static analysis.

Divisions by zero are not portable, however. In languages like C, they are out of spec and hence their behavior can vary from one system to another. While on
Linux x86 systems, an attempted division by zero generates a \texttt{SIGFPE} signal, this is not necessarily the case for ARM Linux. Most ARM processors do not
implement the \texttt{UDIV} and \texttt{SDIV} instructions intended for integer division, emulating these in software instead. On such processors, trying to
divide by zero with the \texttt{UDIV} instruction would not result in a \texttt{SIGFPE}, but rather in a \texttt{SIGILL}, as the instruction is not even
recognized. On other ARM systems (or architectures such as MIPS), no fault or signal is even generated, and the division by zero simply returns a zero.

\subsubsection{Segmentation Faults}

A segmentation fault occurs when upon an invalid memory access. The access might be invalid because the address does not refer to mapped memory at all, such as
a \texttt{NULL} pointer dereference; or because the process lacks the permissions required for the specific access, such as writing to read-only memory, or
trying to execute kernel code. On Linux, a segmentation fault results in a \texttt{SIGSEGV} signal.

We can intentionally generate segmentation faults by inserting instructions that makes invalid memory accesses. The layout of the address space is not very
predictable, however. It contains two fixed areas that we can choose addresses from: the NULL page, and kernel space. Even if the kernel is running with Kernel
Page Table Isolation (to protect against Meltdown attacks), accessing locations in the then unmapped kernel memory still results in a \texttt{SIGSEGV}. The
kernel's exact address range varies between architectures, and some pages are explicitly mapped to be accessible from user space such as the \texttt{vectors}
page on ARMv7 and the \texttt{vsyscall} page on x86. As their addresses are fixed, they pose no real problems. 

On ARM, we have two choices for the specific faulting instructions: load/store instructions and indirect control-flow instructions. At the moment of the
attempted memory access or control transfer, their base register and target address register contain the kernel-space address, which can be computed in an
obfuscated manner to make it less conspicuous. Many more options to generate segmentation faults exist, and the options for which the specific fault address can
easily encode information include the following: writing to the binary's read-only data segment, branching to the binary's data segment, branching to the stack,
etc. On x86, equivalent instructions exist that can be used to trigger segmentation faults.

\subsubsection{Unaligned Memory Accesses}

Even if an instruction accesses a valid address, it can still fault when the address is not alignment. Some ARM processors do not allow any unaligned memory
access. Some only disallow it for certain instructions. Similarly, some x86 instructions also require a specific alignment. In those cases, unaligned memory
accesses are regarded as a software fault, and on Linux result in a \texttt{SIGBUS} signal.

As the instructions that can trigger a \texttt{SIGBUS} vary across ARM processors, there is a clear portability issue. We can sidestep this issue by only
inserting instructions for which unaligned memory accesses are illegal on every ARM processor, such as \texttt{LDREX}, but that is not stealthy at all.
Moreover, a more problematic portability issue is that the Linux kernel on ARM is often configured to hide unaligned memory accesses from user processes. In
that case, the kernel handles unaligned memory accesses without even sending a \texttt{SIGBUS} signal. 

\subsubsection{Actual Implementation}

From our investigation we concluded that triggering segmentation faults is the most platform-independent signaling method. It can be made to work on all ARM
devices, and it is also portable to x86 platforms. In general, we suspect significantly less portability concerns on the x86 platform. The two signaling methods
we implemented in our prototype reflect our conclusions. Both methods use segmentation faults: one uses loads and stores, the other uses indirect branches.

\subsection{Switch Verification}
\label{sec:switch}

Upon invocation, the mini-debugger has to determine whether it is dealing with an actual invocation or a genuine signal caused by original application behavior.
We therefore use a whitelist. For all switches, coming both protected application as self-debugger, this whitelist contains the addresses of all potential
faulting instructions. The addresses in this whitelist should of course be obfuscated. Upon invocation, the mini-debugger verifies the faulting instruction is
whitelist. If not, the signal is passed on to the exception-throwing process.


%% file: evaluation.tex
\section{Evaluation}
\label{sec:evaluation}

We evaluated and tested our technique on an Arndale ARMv7 Board that runs Ubuntu 15.04, featuring Linux 4.3.0. In our tool chain, we used GCC 4.8.1 and GNU
binutils 2.23, the latest versions for which we have the necessary patches to support our binary rewriter. From version 4.4 onwards, Android features Linux
kernels recent enough for our improved mini-debugger, but we lack a patched Android compiler tool chain for such recent versions. We therefore did not evaluate
our technique on Android. The lack of recent patches is an engineering issue that results from a lack of resources to port and maintain patches forward, not a
fundamental issue. All patches are available online, and all our prototype source code will be released upon publication of this paper.

\subsection{Performance Overhead}

\begin{table}[t]
\centering
\caption{Overhead of individual self-debugging aspects}
\begin{tabular}{|l|c|}
\hline
Aspect & Execution Time\\
\hline
Initialization &  16 ms\\
Memory Read & 4.4 $\mu$s\\
Memory Write & 4.8 $\mu$s\\
Switch SIGTRAP & 7.9 ms\\
Switch SIGSEGV (RW) & 0.065 ms\\
Switch SIGSEGV (X) & 0.060 ms\\
\hline
\end{tabular}
\label{table:overhead}
\end{table}

It is rather meaningless to present a concrete performance overhead for an application protected by our improved self-debugging technique. When protecting their
application, developers choose which code fragments to migrate to the self-debugger in function of the specific, sensitive code they want to protect. To thwart
attackers, they should ensure plenty of control flow switches occur in or near the sensitive code. The overhead for a specific application depends by and large
on the fragments being migrated and on which code paths these fragments lie. Profiling can be used to select code fragments that are not too hot, nor too cold,
but the exact performance overhead is context-dependent. On top of that, we can transform hot code to create colder code that is more suitable to being
migrated, through loop tiling for example.

Still, we can measure the impact of several operations, such as its initialization, memory operations performed from a migrated code fragment, and switches to
and from such a fragment. To that extent, we used micro-benchmarks that were completely dominated by the aspect to measure, compiled these benchmarks at
\texttt{-O2} and employed \texttt{perf}. Our results can be seen in Table~\ref{table:overhead}. We measured the execution time for: the self-debugging
initialization routine; memory reads or writes from the self-debugger, using the mini-debugger's memory support; and control flow switches, using the different
signaling methods we implemented. The memory reads and writes correspond to a single \texttt{LDR} or \texttt{STR} instruction. Two switches occur when executing
a migrated code fragment in the self-debugger: one switch to the self-debugger, and one switch back the application.

We measured the \texttt{SIGTRAP} signaling method employed in our previous implementation as a reference, and came to the conclusion that it is two orders of
magnitude slower as the stealthier signaling methods we implemented. The reason for this is that the specific Linux kernel code path triggered by a
\texttt{BKPT} instruction performs some I/O, slowing it down~\cite{bkpt_mystery}.

\subsection{Security Evaluation}

\begin{table}[t]
\centering
\caption{Results from Binary Ninja analysis}
\resizebox{\linewidth}{!}{
\begin{tabular}{|c|cc|cc|cc|}
\hline
  \textbf{\# Code}   & \multicolumn{2}{c|}{\textbf{Migrated Targets}} & \multicolumn{2}{c|}{\textbf{Non-migrated Targets}} & \multicolumn{2}{c|}{\textbf{Faulting Instructions}} \\
  \textbf{Fragments} & Disassembled & Total                       & Disassembled & Total                           & Value Found & Total\\
\hline
  3  & 2  & 3  & 1  & 5  & 3  & 9  \\
  5  & 2  & 5  & 4  & 7  & 5  & 14 \\
  7  & 2  & 7  & 6  & 9  & 4  & 19 \\
  10 & 2  & 10 & 4  & 12 & 7  & 25 \\
  13 & 3  & 13 & 8  & 15 & 6  & 32 \\
  15 & 4  & 15 & 10 & 17 & 10 & 36 \\
\hline
\end{tabular}
}
\label{table:binary_ninja}
\end{table}

We evaluate how our improved design thwarts several approaches an attacker could take to defeat our original design, and conclude this section by considering
the attack approaches that remain, as well as the orthogonal defenses that can improve the resilience of our technique.

In our original design, an attacker could attach their own debugger to the self-debugger. In our improved design, such a debugger can no longer attach to either
the self-debugger or the protected application. We verified this in real life, but also investigated an alternative attack that, with some extra work, still
allows an attacker to attach a hostile debugger to either process, although only temporarily. We explain this attack with the example goal of attaching to the
protected application. To achieve this goal, the attacker needs to simultaneously terminate the self-debugger through a \texttt{SIGKILL}, and pause the
protected application through a \texttt{SIGSTOP}. The latter signal is necessary to prevent the protected application from continuing its execution,
encountering a mini-debugger invocation masquerading as a software fault, and dying. The attacker can subsequently attach their debugger to the application,
continue it, and analyze or tamper with its execution until it encounters the next software fault. To perform this attack the \verb!PTRACE_O_EXITKILL! option
first has to be disabled, however. If it were not, the \texttt{SIGKILL} signal would result in the death of both processes.

After tampering with the binary to disable \verb!PTRACE_O_EXITKILL!, an attacker can thus temporarily attach to either process and observe or tamper with the
execution of application code (migrated or otherwise). To stay attached more permanently, however, the attacker has to reverse the transformations our technique
employs to migrate code and invoke the mini-debugger. In our previous design, such invocations could easily be recognized statically through their use of
\texttt{BKPT} instructions. In our improved design, this is no longer the case as we reuse application code to trigger software faults. If an attacker were to
invest extra time and effort, a more advanced static analysis might be able to find the faulting instructions, however.

To evaluate how easy it was to find faulting instructions statically, we set up an experiment using the open-source compression tool bzip2, compiled at
\texttt{-Os}. We successively migrated more code fragments to the self-debugger, and analyzed the resulting binary using Binary Ninja~\cite{ninja}. The reason
we use Binary Ninja as opposed to other disassemblers (such as IDA Pro) is that it contains more advanced---and for this experiment, more relevant---features
such as value-set analysis (VSA). Table~\ref{table:binary_ninja} contains the results of Binary Ninja's analysis. The only protection we applied to bzip2 was
our own self-debugging technique; we did not use any obfuscations. Even then, VSA found the faulting address as a possible value for only roughly a third of all
faulting instructions. Binary Ninja does not explicitly flag these values as causing software faults, but an attacker could write an analysis to do so. On top
of that, Binary Ninja's recursive descent no longer disassembles all the targets of control transfers, regarding them as data instead. The control transfers
from the mini-debugger to these targets happen through a complex indirection, and can not be followed. Some targets have more direct incoming control flow that
causes them to be disassembled, however, or are recognized as code without incoming control flow by heuristics. Our stealthier interfaces are thus rather
effective in hiding mini-debugger invocations, but could be even more effective if augmented with more advanced data obfuscations.

If attackers cannot locate and reverse our transformations statically, they have to do so dynamically. Our technique's entire purpose is to hinder dynamic
analysis using debuggers, however. Reversing our transformations one software fault at a time would come at a considerable cost. Our technique can be made even
stronger by applying orthogonal defenses such as (data) obfuscations and integrity checks, particularly on the initialization routine where the
\verb!PTRACE_O_EXITKILL! option is enabled. Other, less convenient, potential approaches still left open to an attack are emulation, which we consider out of
scope; and some sort of emulation/interposing of the \texttt{ptrace} API, which would again require considerable effort.


%% file: conclusions.tex
\section{Conclusions}
\label{sec:conclusions}

We improved the state-of-the-art anti-debugging technique of self-debugging by introducing reciprocal debugging and by making the interfaces to the
self-debugger stealthier. The goal of these improvements was to make it harder for attackers to defeat the protection. Our evaluation confirms that this goal
was reached. 
